\documentclass[twocolumn]{aastex631}

\usepackage{textcomp, gensymb}
\usepackage{graphicx}
\usepackage{natbib}
\usepackage[normalem]{ulem}

\begin{document}

\title{Optical and Near-Infrared Contemporaneous Polarimetry of C/2023 A3 (Tsuchinshan-ATLAS)}

\author[0000-0002-8244-4603]{Bumhoo Lim}
\affiliation{Department of Physics and Astronomy, Seoul National University, 1 Gwanak-ro, Gwanak-gu, Seoul 08826, Korea}
\affiliation{SNU Astronomy Research Center, Department of Physics and Astronomy, Seoul National University, 1 Gwanak-ro, Gwanak-gu, Seoul 08826, Republic of Korea}

\author[0000-0002-7332-2479]{Masateru Ishiguro}
\affiliation{Department of Physics and Astronomy, Seoul National University, 1 Gwanak-ro, Gwanak-gu, Seoul 08826, Korea}
\affiliation{SNU Astronomy Research Center, Department of Physics and Astronomy, Seoul National University, 1 Gwanak-ro, Gwanak-gu, Seoul 08826, Republic of Korea}

\author[0000-0002-2928-8306]{Jun Takahashi}
\affiliation{Center for Astronomy, University of Hyogo, 407-2 Nishigaichi, Sayo, Hyogo 679-5313, Japan}

\author[0000-0001-6156-238X]{Hiroshi Akitakya}
\affiliation{Astronomy Research Center, Chiba Institute of Technology, 2-17-1 Tsudanuma, Narashino, Chiba 275-0016, Japan}
\affiliation{Planetary Exploration Research Center, Chiba Institute of Technology, 2-17-1 Tsudanuma, Narashino, Chiba 275-0016, Japan}
\affiliation{Hiroshima Astrophysical Science Center, Hiroshima University, 1-3-1 Kagamiyama, Higashi-Hiroshima, Hiroshima 739-8526, Japan}

\author[0000-0002-3291-4056]{Jooyeon Geem}
\affiliation{Department of Physics and Astronomy, Seoul National University, 1 Gwanak-ro, Gwanak-gu, Seoul 08826, Korea}
\affiliation{SNU Astronomy Research Center, Department of Physics and Astronomy, Seoul National University, 1 Gwanak-ro, Gwanak-gu, Seoul 08826, Republic of Korea}

\author[0000-0002-2618-1124]{Yoonsoo P. Bach}
\affiliation{Korea Astronomy and Space Science Institute, Daejeon 305-348, Korea}

\author[0000-0002-0460-7550]{Sunho Jin}
\affiliation{Department of Physics and Astronomy, Seoul National University, 1 Gwanak-ro, Gwanak-gu, Seoul 08826, Korea}
\affiliation{SNU Astronomy Research Center, Department of Physics and Astronomy, Seoul National University, 1 Gwanak-ro, Gwanak-gu, Seoul 08826, Republic of Korea}

\author[0009-0004-9591-8646]{Hangbin Jo}
\affiliation{Department of Physics and Astronomy, Seoul National University, 1 Gwanak-ro, Gwanak-gu, Seoul 08826, Korea}
\affiliation{SNU Astronomy Research Center, Department of Physics and Astronomy, Seoul National University, 1 Gwanak-ro, Gwanak-gu, Seoul 08826, Republic of Korea}

\author{Seungwon Choi}
\affiliation{Department of Physics and Astronomy, Seoul National University, 1 Gwanak-ro, Gwanak-gu, Seoul 08826, Korea}
\affiliation{SNU Astronomy Research Center, Department of Physics and Astronomy, Seoul National University, 1 Gwanak-ro, Gwanak-gu, Seoul 08826, Republic of Korea}

\author{Jinguk Seo}
\affiliation{Department of Physics and Astronomy, Seoul National University, 1 Gwanak-ro, Gwanak-gu, Seoul 08826, Korea}
\affiliation{SNU Astronomy Research Center, Department of Physics and Astronomy, Seoul National University, 1 Gwanak-ro, Gwanak-gu, Seoul 08826, Republic of Korea}

\author[0000-0001-6099-9539]{Koji S. Kawabata}
\affiliation{Hiroshima Astrophysical Science Center, Hiroshima University, 1-3-1 Kagamiyama, Higashi-Hiroshima, Hiroshima 739-8526, Japan}

\author{Tomoya Hori}
\affiliation{Graduate School of Advanced Science and Engineering, Hiroshima University, 1-3-1 Kagamiyama, Higashi-Hiroshima, Hiroshima 739-8526, Japan}

\author{Tetsuharu Maruta}
\affiliation{Graduate School of Advanced Science and Engineering, Hiroshima University, 1-3-1 Kagamiyama, Higashi-Hiroshima, Hiroshima 739-8526, Japan}

\author[0000-0002-8537-6714]{Myungshin Im}
\affiliation{Department of Physics and Astronomy, Seoul National University, 1 Gwanak-ro, Gwanak-gu, Seoul 08826, Korea}
\affiliation{SNU Astronomy Research Center, Department of Physics and Astronomy, Seoul National University, 1 Gwanak-ro, Gwanak-gu, Seoul 08826, Republic of Korea}


\correspondingauthor{Masateru Ishiguro}
\email{bumhoo7@snu.ac.kr, ishiguro@snu.ac.kr}

\begin{abstract}
We conducted contemporaneous optical and near-infrared polarimetric and spectroscopic observations of C/2023 A3 (Tsuchinshan-ATLAS, hereafter T-A) from 2024 October 16 to December 17, covering a wide range of phase angles (20--123\degr) and wavelength (0.5--2.3 \micron). We paid special attention to gas contamination in the dust polarization using these data. As a result, we find the maximum polarization degree $P_\mathrm{max}=31.21\pm0.05$ \%, $33.52\pm0.06$ \%, $35.12\pm0.01$ \%, $37.57\pm0.01$ \%, and $35.35\pm0.01$ \% in the Rc-, Ic-, J-, H-, and Ks-bands, respectively. Although dust polarization shows a red slope at shorter wavelengths and can peak around 1.6 \micron, the phase angle at which maximum polarization occurs exhibits less dependence on wavelength ($\alpha_\mathrm{max}\sim90\degr$ -- $95\degr$). Although only a few historically bright comets, such as West, Halley, and Hale-Bopp, have undergone such extensive dust-polarization observations, our measurements are generally consistent with those of two comets that possibly originated from the Oort Cloud (West and Halley). From these results, we conjecture that the optical properties and growth processes of dust in the presolar nebula, which formed these cometary nuclei, were likely uniform.
\end{abstract}


\section{Introduction} \label{sec:introduction}

Among the Small Solar System Bodies (SSSBs), comets are considered to be the most primitive objects that have preserved information about the formation epoch of the Solar System approximately 4.6 billion years ago. These comets are broadly classified into long-period comets or non-periodic comets originating from the Oort Cloud (Oort Cloud comets, OCCs) and short-period comets with small orbital inclinations originating from the Kuiper Belt (Ecliptic comets, ECs). OCCs are believed to have formed in gas planet-forming regions \citep{2000Icar..145..580F}, while ECs formed outside of planet-forming regions \citep{1994Icar..108...18L}. Except for some comets that evolved under solar heating or large icy bodies that might have held an internal heat source \citep{2002EM&P...89...27P}, small comets would be subject to a less thermal alteration process. Recent in-situ observations by space probes have dramatically enhanced our pictures of ECs.
However, collisional and thermal evolutions of the ECs cannot be ignored, which may have obscured primordial information. In contrast, given the enormous volume, OCCs are believed to be objects that have experienced few collisions \citep{2003Natur.424..639S}. Therefore, OCCs are unique research targets that retain a great deal of primordial information about regions where gas giants formed.

Reviews of SSSB polarimetry, including comets, are given in \citet{2015psps.book..379K,2024A&ARv..32....7B}. In particular, comet polarimetry provides information such as optical properties, monomer size, and porosity of dust grains in the comae \citep[see, e.g.,][]{2006A&A...449.1243K}. At optical wavelengths, comet polarimetry was performed for many comets (as of March 2025, 95 comets were archived on the NASA Planetary Data System, \citealt{2017PDSS..271.....K}). Initially, it was suggested that there were two main types: high-polarized and low-polarized groups \citep{1996A&A...313..327L}. Since the low-polarization group indicated strong gas emissions (lower polarization than dust) at the optical wavelength, the dichotomy of the polarimetric groups may result from contamination by molecular fluorescence emissions \citep{2001SoSyR..35..480K, 2005A&A...441..773J}, although recent studies suggest that low polarization in comets could be an intrinsic property of the dust particles \citep{2020Icar..33613453Z, 2024EPSC...17..343Z}. On the other hand, because there are no prominent emissions in the near-infrared wavelength (NIR), NIR polarimetry captures less-contaminated signatures of dust scattering, although the influence of thermal radiation is not negligible for comets at a small distance from the Sun \citep{1978PASJ...30..161O}. NIR polarimetry has been performed for a dozen comets at limited phase angles and limited wavelengths. \citet{2004AJ....127.2398K} obtained NIR polarimetric data only for K- or HK-bands, and \citet{2019A&A...629A.121K} only for JHK-bands (see Section \ref{sec:discussion}). Only three comets (West, 1P/Halley, and Hale-Bopp) have polarization data with the same phase angles over a wide wavelength range in the optical and NIR \citep{1978PASJ...30..161O, 1987A&A...187..621B, 1987A&A...187..689K, 1997EM&P...78..353H, 2006JQSRT.100..179K}. Here, we emphasize that previous NIR polarimetry has been conducted by acquiring data at different phase angles, making it difficult to compare the physical properties of individual comets.

Considering this situation of comet polarimetry, we conducted coordinated observations of a bright comet C/2023 A3 (Tsuchinshan-ATLAS, hereafter T-A) to obtain nearly simultaneous data in the optical and NIR at a wide range of phase angles ($\alpha$=20--123\degr). This comet is one of the brightest ones of this century, and it was observable with the naked eye from the northern hemisphere, making it possible to conduct this coordinated monitoring observation using a network of small- and medium-aperture telescopes. The observations were made at four different observatories, all of which were timed to be as close as possible to each other. In addition to polarimetry, we conducted spectroscopy to examine the influence of molecular emissions. 

In Section \ref{sec:observations}, we describe the outlines of the observations and data analyses, and in Section \ref{sec:results}, we present the observation results. In Section \ref{sec:discussion}, we compare our results with those of the three great comets (West, 1P/Halley, and Hale-Bopp) and consider their dusty nature.

\section{Observations and Data Analyses} \label{sec:observations}

The details of the observations are summarized in Table \ref{tab:observation_log}. The optical imaging polarimetry was conducted using the Seoul National University (SNU) Quadruple Imaging Device for POLarimetry (SQUIDPOL) attached to the Cassegrain focus of the 0.6-m telescope at Pyeongchang Observatory of SNU. We chose the standard Bessel filters for the Rc- and Ic-bands. Near-infrared imaging polarimetry was performed using the polarimetric mode of the Nishiharima Infrared Camera (NIC) attached to the 2.0-m Nayuta telescope at the Nishi-Harima Astronomical Observatory (NHAO), University of Hyogo. NIC is capable of simultaneous J-, H-, and Ks-bands imaging, including imaging polarimetry \citep{2019StGal...2....3T}. For optical-NIR spectropolarimetry, we used the Hiroshima Optical and Near-InfraRed camera (HONIR) attached to the Cassegrain focus of the 1.5-m Kanata telescope at Higashi-Hiroshima Observatory (HHO), Hiroshima University \citep{2014SPIE.9147E..4OA}. We chose an IR long grism and 200 \micron\ slit (2.2\arcsec). This combination enables us to take the data at 0.5--1.0 \micron\ for the optical channel and 1.4--2.4 \micron\ for the infrared channel at the same time. In this study, only the optical channel data from HONIR are presented due to significant noise in the infrared data. All the polarimeters in our observations were equipped with beam splitter-type polarimetric elements to minimize uncertainties such as variable weather and imperfect flat field correction. 

In addition to these polarimetric observations, we performed long-slit spectroscopy using the 1.0-m telescope at the SNU Astronomical Observatory (SAO) on the SNU Gwanak campus and a ready-made LISA spectrograph from Shelyak Instruments. We used a 23 \micron\ (0.79\arcsec) slit, which results in the spectral resolution of $R\sim1000$.

Although these observations were coordinated to obtain data as coincidentally as possible, some data were not being conducted at perfectly the same time for some reason (e.g., different weather conditions, elevation and sunset time, or arrangements with other observation programs). In every dataset, we utilized the data with good weather and seeing ($<$2.5\arcsec) conditions. We analyzed these data in a standard manner, as seen in previous research \citep{2014SPIE.9147E..4OA,2019StGal...2....3T,2022StGal...5....4B}. The outline includes preprocessing (flat, bias, and dark corrections), extraction of target signals, correction for instrumental polarization and polarization efficiency, derivation of the normalized Stokes parameters ($Q/I$ and $U/I$) and the polarization degrees perpendicular to the comet's scattering plane ($P_\mathrm{r})$ for polarimetric data, and wavelength and flux calibrations for spectral data. 

\begin{deluxetable*}{ccccccccccc}[!htbp]
\tablecaption{Observational log. \label{tab:observation_log}}
\tablehead{
\colhead{UT Date} & \colhead{Mode} & \colhead{Filter/Wavelength} & \colhead{Exptime [s]} & 
\colhead{N} & \colhead{$r_{\rm{h}}$ [au]} & \colhead{$\Delta$ [au]} & 
\colhead{$\alpha$ [$\degr$]} & \colhead{Airmass} & \colhead{Seeing [\arcsec]}
\\
(1) & (2) & (3) & (4) & (5) & (6) & (7) & (8) & (9) & (10)}
\startdata
\multicolumn{9}{l}{SNU/SQUIDPOL} \\
        2024-10-16.41  & impol & Rc, Ic               & 30      & 4, 11   & 0.63 & 0.51  &        123.18  & 9.59  & 2.5  \\
        2024-10-17.42  & impol & Rc, Ic               & 30      & 7, 7    & 0.64 & 0.53  &        116.50  & 10.62 & 2.0  \\
        2024-10-20.42  & impol & Rc, Ic               & 30      & 14, 5   & 0.70 & 0.60  &         99.67  & 6.45  & 2.2  \\
\textbf{2024-10-23.42} & impol & Rc, Ic               & 30      & 9, 3    & 0.76 & 0.69  & \textbf{86.74} & 3.29  & 2.2  \\
        2024-10-24.43  & impol & Rc, Ic               & 180, 60 & 18, 18  & 0.78 & 0.72  &         83.00  & 6.34  & 2.1  \\
\textbf{2024-10-30.44} & impol & Rc, Ic               & 60      & 25, 25  & 0.90 & 0.92  & \textbf{66.14} & 6.36  & 1.9  \\
        2024-10-31.43  & impol & Rc, Ic               & 180, 60 & 28, 18  & 0.92 & 0.96  &         63.96  & 6.11  & 2.2  \\
\textbf{2024-11-12.42} & impol & Rc, Ic               & 180     & 10, 9   & 1.14 & 1.37  & \textbf{45.25} & 3.81  & 2.3  \\
        2024-11-19.43  & impol & Rc                   & 180     & 1       & 1.27 & 1.60  &         38.08  & 3.97  & 1.9  \\
        2024-12-17.42  & impol & Rc                   & 120     & 4       & 1.75 & 2.40  &         20.61  & 11.57 & 1.8  \\
\hline
\multicolumn{9}{l}{NHAO/NIC} \\
\textbf{2024-10-23.42} & impol & J, H, Ks             & 30, 60  & 5, 6, 3 & 0.76 & 0.69  & \textbf{86.71} & 2.58  & 2.3  \\
        2024-10-25.39  & impol & J, H, Ks             & 90      & 5, 4, 3 & 0.80 & 0.75  &         79.78  & 1.80  & 1.9  \\
\textbf{2024-10-30.40} & impol & J, H, Ks             & 90      & 7, 8, 8 & 0.90 & 0.92  & \textbf{66.22} & 1.85  & 2.1  \\
        2024-11-11.40  & impol & J, H, Ks             & 120     & 6, 7, 6 & 1.12 & 1.34  &         46.46  & 1.95  & 2.2  \\
\textbf{2024-11-12.39} & impol & J, H, Ks             & 120     & 6, 6, 6 & 1.14 & 1.37  & \textbf{45.28} & 1.82  & 2.5  \\
\hline
\multicolumn{9}{l}{HHO/HONIR} \\
        2024-10-20.41  & sppol & OPT\tablenotemark{a} & 80      & 3       & 0.70 & 0.60  &         99.76  & 1.75  & 2.2  \\
\textbf{2024-10-30.44} & sppol & OPT                  & 80      & 6       & 0.90 & 0.92  & \textbf{66.13} & 1.67  & 2.4  \\
\textbf{2024-11-12.42} & sppol & OPT                  & 80      & 8       & 1.14 & 1.37  & \textbf{45.25} & 1.77  & 2.4  \\
\hline
\multicolumn{9}{l}{SAO/LISA} \\
\textbf{2024-10-30.42} & spec  & VIS\tablenotemark{b} & 600     & 5       & 0.90 & 0.92  & \textbf{66.17} & 1.80   & 2.2  \\
        2024-10-31.41  & spec  & VIS                  & 600     & 5       & 0.91 & 0.96  &         64.01  & 1.72   & 1.9  \\
\enddata
\tablenotetext{a}{Optical grism covering the wavelength range of $\lambda=0.58$--$1.0\ \micron$ \citep{2014SPIE.9147E..4OA}.}
\tablenotetext{b}{Visible domain covering the wavelength range of $\lambda=0.4$--$0.7\ \micron$.}
\tablecomments{
(1) Midpoint of the observation periode. (2) Observation mode, selected from imaging polarimetry (impol), spectropolarimetry (sppol), or spectroscopy (spec). (3) Filters used in each observation. (See Table \ref{tab:polparams} for detailed filter information.) (4) Individual exposure time. (5) Number of data sets used for each filter. (6) Median heliocentric distance. (7) Median geocentric distance. (8) Median phase angle. (9) Median airmass. (10) Typical seeing scale (FWHM). The ephemerides are obtained from the web-based JPL Horizon system (\url{http://ssd.jpl.nasa.gov/?horizons}). \textbf{Bolded dates} indicate the coordinated observations across multiple observatories.
}
\end{deluxetable*}

\section{Results} \label{sec:results}

In this section, we show our findings on the phase angle dependency in Section \ref{subsec:phaseangle}, evaluation of the effects of fluorescence emission in Section \ref{subsec:gas}, and the wavelength dependence in Section \ref{subsec:spectral}.

\subsection{Phase angle dependence}
\label{subsec:phaseangle}

\begin{figure*}[!htbp]
    \centering
    \includegraphics[width=1.0\linewidth]{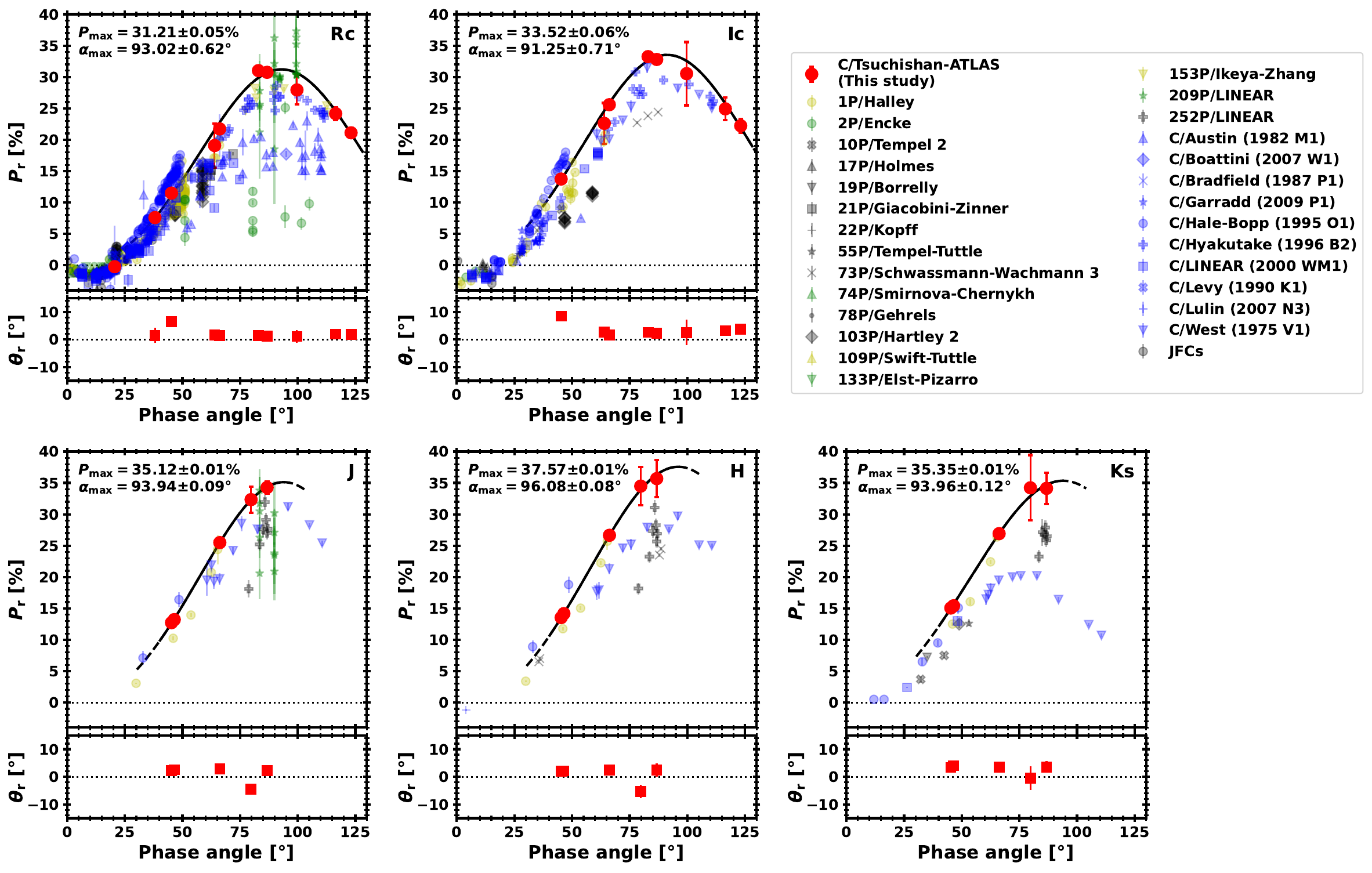}
    \caption{Polarization degree ($P_{\mathrm{r}}$) and position angle of the polarization vector ($\theta_{\mathrm{r}}$) for phase angle at five bands (Rc, Ic, J, H, and Ks). The red color indicates data points from this study, whereas black, blue, green, and yellow indicate that from other comets in similar bands, which is retrieved from DataBase of Comet Polarimetry archive \citep[DBCP;][]{2017PDSS..271.....K}, except 252P/LINEAR \citep{2019A&A...629A.121K}. The solid lines show the fitting curve of data points with Equation (\ref{eq:ppcfit}), while excluding zero-polarization data in Rc-band near $\alpha\sim20\degr$. The wavelength ranges of each band are as follows: Rc (620--730 nm), Ic (730--1050 nm), J (1050--1400 nm), H (1400--1800 nm), and Ks (1800--2500 nm).}
    \label{fig:ppc}
\end{figure*}

Table \ref{tab:impol} summarizes the results of linear polarimetry. To obtain these values, we derived source fluxes on the ordinary and extraordinary components within the aperture of 10,000--15,000 km ($\sim$13\arcsec) from the brightest locus of the coma (where the nucleus would exist) and subtracted the sky region by setting the annulus of 30,000--62,000 km ($\sim$1\arcmin) from the nucleus. The aperture size was selected after examining the radial profile of the coma brightness and the polarization degree to ensure that it encompassed the entire dust coma signal. In general, the scattered sunlight from SSSBs (including comet comae) indicates a strong dependence on the solar phase angle (the Sun object-observer angle, $\alpha$). Following the convention of SSSBs polarimetry, in Figure \ref{fig:ppc}, we plot the polarization degrees with respect to the scattering plane in terms of $\alpha$ at each band and compare them with other comets. This plot is known as the polarization phase curve (PPC). At first glance, the PPC plots indicate that $P_\mathrm{r}$ increase almost linearly until $\alpha\sim80$\degr, have maximums around $\alpha\sim90\degr$, and decrease beyond the phase angle. In the Rc-band, the polarization degree reaches to zero ($P_\mathrm{r}=0\degr$) at an inversion angle $\alpha_\mathrm{inv}\sim20\degr$, implying the presence of a negative branch ($P_\mathrm{r}<0\degr$). This trend is typical in polarimetry of SSSBs and other Solar System objects, including the Moon.

To determine the maximum values of polarization degree ($P_{\rm{max}}$) and the corresponding phase angles ($\alpha_{\rm{max}}$), we utilized an empirical function given in \citet{2005SoSyR..39...45K},

\begin{equation}\label{eq:ppcfit}
    P_\mathrm{r}(\alpha)=\frac{(\sin^2(\alpha-\Delta\alpha))^W}{1+\cos^2(\alpha-\Delta\alpha)+D}~,
\end{equation}

\noindent where the three free parameters, $\Delta\alpha$, $W$, and $D$, characterize a shift in $\alpha_{\rm{max}}$ from $90\degr$, the width of the positive branch and the degree of depolarization, respectively. Since Equation (\ref{eq:ppcfit}) was constructed to fit the positive branch ($P_\mathrm{r}>0\degr$) of PPC in the lunar regolith \citep{2005SoSyR..39...45K}, we excluded the zero-polarization data in our PPC fitting. We found that this empirical curve fits well with our data, resulting in a reduced chi-square value smaller than 3 ($\chi_\nu<3$) for all the five-band data.

In Figure \ref{fig:ppc}, we draw the fitting results with solid lines. We find that the polarization degree has a maximum of $P_{\rm{max}}=31.21\pm0.05\%$ at $\alpha$=$\alpha_{\rm{max}}=93.02\pm0.62\degr$ in the Rc-band and $P_{\rm{max}}$ increases as the observed wavelength increases. In contrast, the $\alpha_{\rm{max}}$ values (which is consistent with other comets in the Rc-band, \citealt{2024A&ARv..32....7B}) are nearly constant regardless of the wavelengths. 

In the optical channels, the T-A's PPC at large phase angles ($\alpha\gtrsim 60$\degr) shows good agreement with many comets, including OCCs (C/Hyakutake and C/West) and Halley-type (1P/Halley)  \citep{1996A&A...313..327L}. However, the PPC at large phase angles significantly exceeds all five other comets in the NIR (West, 1P/Halley, 73P/S-W, 209P/LINEAR, and 252P/LINEAR).

\subsection{Influence of fluorescence emissions}
\label{subsec:gas}
\begin{figure*}[!htbp]
    \centering
    \includegraphics[width=1.0\linewidth]{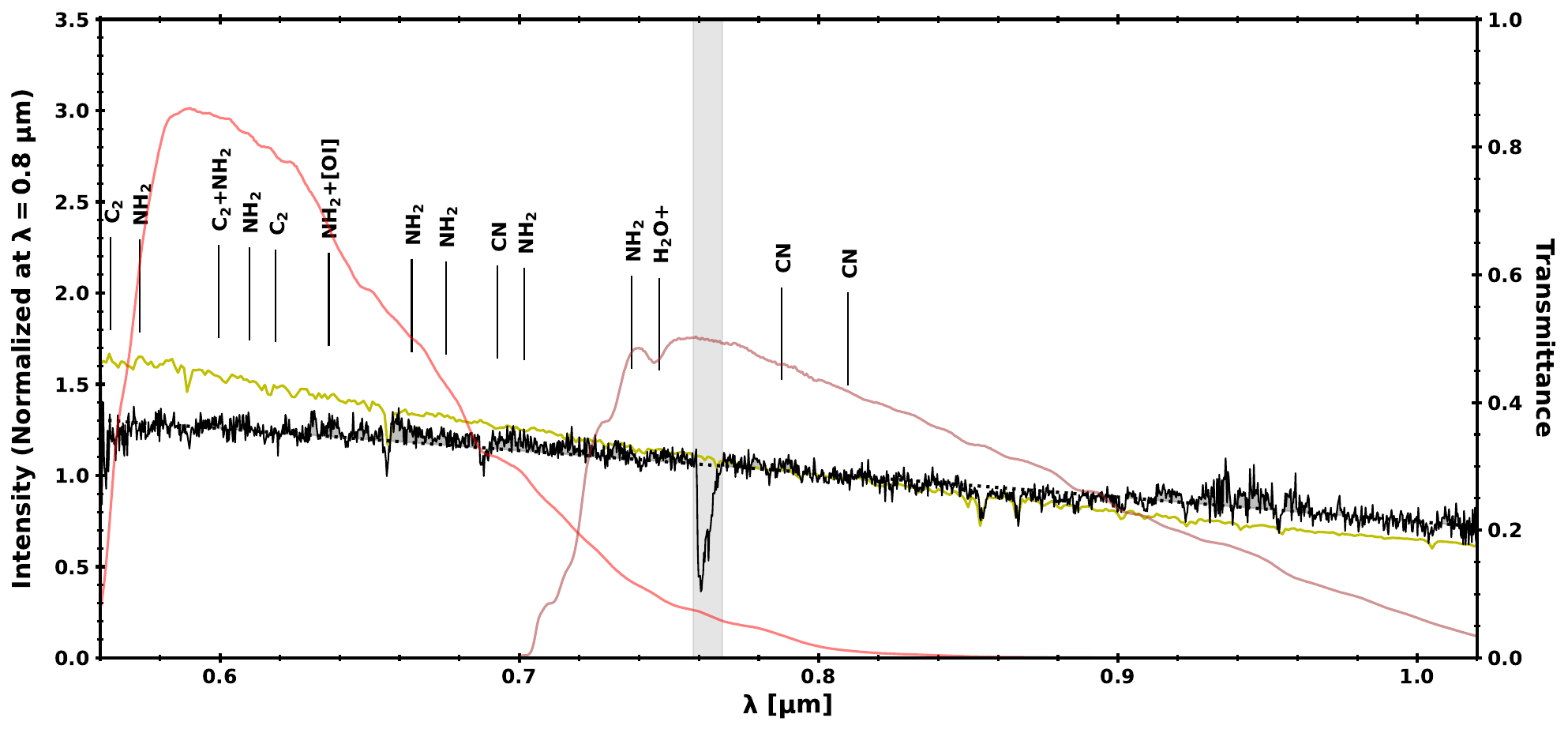}
    \caption{HONIR median-combined spectrum of T-A observed on November 12, 2024. The dotted line shows the secondary polynomial fit of the continuum, and the yellow color shows the extraterrestrial solar spectrum. Both T-A and solar spectrum are normalized with the flux at $\lambda=0.8\ \rm{\mu}$m. The red and brown colors show the transmittance of Rc- and Ic-bands in SQUIDPOL. Typical molecular emissions of comets are noted with the species \citep{1996AJ....112.1197B}. The negative spike near $\lambda=0.76\ \rm{\mu}$m is the artifact during flux calibration due to the imperfect correction of telluric absorption.}
    \label{fig:spectrum}
\end{figure*}

Although we show our polarimetric result above, it is important to check the influence of molecular fluorescence emission. As we mentioned in Section \ref{sec:introduction}, polarimetric signals of comet comae are often contaminated by gas emission, making it difficult to extract the real dust characteristic \citep{2001SoSyR..35..480K, 2005A&A...441..773J}. In particular, there are strong molecular fluorescence emission lines in the optical ($\lambda\lesssim1\ \rm{\mu m}$), which reduce the polarization degree. To examine this effect, we performed spectroscopy and spectropolarimetry. Figure \ref{fig:spectrum} compares the spectrum of the near-nuclear coma region (4,500 km $\times$ 1,500 km, oriented along North-South and East-West directions) with a reference solar spectrum \citep{2004SoEn...76..423G}. We find that the coma has a spectrum redder than the Sun, while it has no prominent emissions in the Rc- and Ic-bands.

To evaluate this effect more quantitatively, we calculated the ratio of the emission line with respect to the continuum, which is given by,

\begin{equation}\label{eq:gasfrac}
    f_{\rm{gas}} = \frac{\int_{\lambda_1}^{\lambda_2} [I_{\rm{obs}}(\lambda)-I_{\rm{cont}}(\lambda)] \, d\lambda}{\int_{\lambda_1}^{\lambda_2} I_{\rm{obs}}(\lambda) \, d\lambda}~,
\end{equation} 

\noindent where $I_{\rm{obs}}(\lambda)$ and $I_{\rm{cont}}(\lambda)$ are the observed and continuum fluxes at the given wavelength $\lambda$ (i.e., solid and dotted lines in Figure \ref{fig:spectrum}). The wavelength ranges ($\lambda_1$, $\lambda_2$) used for each band were $\lambda_1=0.58\ \rm{\mu m}$ and $\lambda_2=0.76\ \rm{\mu m}$ for the Rc-band and $\lambda_1=0.74\ \rm{\mu m}$ and $\lambda_2=0.88\ \rm{\mu m}$ for the Ic-band. The result yields $f_{\rm{gas,\ Rc}}=0.48 \pm 0.03\ \%$ and $f_{\rm{gas,\ Ic}}=0.17 \pm 0.02\ \%$, respectively. We conducted spectroscopic observations on three separate nights (October 30, 31, and November 12) during one month of coordinated observations (October 16 to November 19). All three datasets show similar trends (Appendix \ref{appendix:spectrum}). Therefore, we conclude that gas contamination in our polarimetry is negligible for the observed optical bands (Rc- and Ic-bands).

\subsection{Wavelength dependence}
\label{subsec:spectral}

\begin{figure*}[!htbp]
    \centering
    \includegraphics[width=1.0\linewidth]{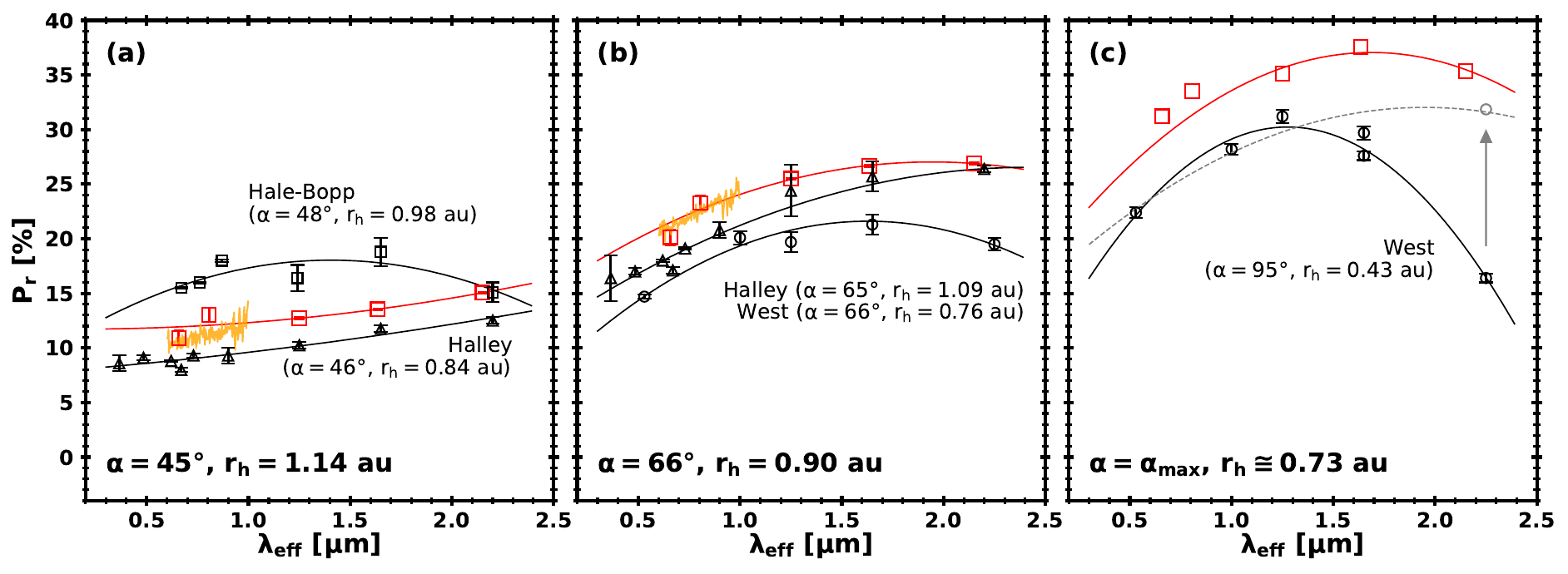}
    \caption{Spectral dependence of polarization degree at three phase angles (a) $\alpha=45\degr$, (b) $\alpha=66\degr$, and (c) $\alpha=\alpha_{\rm{max}}$. In the case of (c), we present the $P_\mathrm{max}$ values from the PPC fitting (Equation (\ref{eq:ppcfit})). The orange spectrum shows the spectropolarimetry data from HONIR at (a) $\alpha=45\degr$ and (b) $\alpha=65\degr$. For comparison, data from comets Hale-Bopp (square), 1P/Halley (triangle), and West (circle) are provided along with phase angles and heliocentric distances at the time of observation. The solid lines indicate secondary polynomial fitting. The gray color in (c) shows the data point and polynomial fitting after correcting the thermal radiation in the K-band ($\lambda=2.2\ \micron$; \citealt{1978PASJ...30..161O}). The polynomial fitted lines are plotted for visualization purposes only.}
    \label{fig:pcolor}
\end{figure*}

Figure \ref{fig:pcolor} shows the spectral dependence of the polarization degree for T-A.
In the plot, we added spectropolarimetric data taken using HONIR (orange color), which is consistent with aperture polarimetry in the Rc- and Ic-bands. Our spectropolarimetry reveals a smooth increase in polarization degree, confirming that gas emissions are not noticeable at these wavelengths (see also Appendix \ref{appendix:sppol}).

We fitted the data points with a second-order polynomial function for visualization purposes (solid black and red lines in Figure \ref{fig:pcolor}). Then, we computed the spectral slopes of the polarization degree ($\% / \micron$), which are given by

\begin{equation}\label{eq:polcolor}
    c_{i,j}:=\frac{P_{\mathrm{r},j}\left(\lambda_j\right)-P_{\mathrm{r},i}\left(\lambda_i\right)}{\lambda_j-\lambda_i}~~,
\end{equation}

\noindent where $P_{\mathrm{r},i}\left(\lambda_i\right)$ and $P_{\mathrm{r},j}\left(\lambda_j\right)$ are polarization degrees measured at effective wavelengths $\lambda_i$ and $\lambda_j$ (with $\lambda_i<\lambda_j$), respectively. We calculated $c_{i,j}$ of T-A in optical ($c_{RH}$) and NIR ($c_{HK}$), where $\lambda_R=0.6\ \micron$, $\lambda_H=1.6\ \micron$, and $\lambda_K=2.2\ \micron$ (i.e., Rc-, H-, and Ks-bands, respectively) at three phase angles: $\alpha = 45\degr$, $\alpha = 66\degr$, and $\alpha = \alpha_{\mathrm{max}}$. The results are as follows: $c_{RH}=2.68\pm0.69\ \% / \micron$ and $c_{HK}=2.96\pm0.23\ \% / \micron$ ($\alpha=45\degr$); $c_{RH}=6.67\pm0.73\ \% / \micron$ and $c_{HK}=0.51\pm0.26\ \% / \micron$ ($\alpha=66\degr$); and $c_{RH}=6.51\ \% / \micron$ and $c_{HK}=-4.33\ \% / \micron$ ($\alpha=\alpha_{\mathrm{max}}$). We find that spectral slopes of the polarization degree (i) are generally positive (i.e., red polarimetric colors), (ii) increase with increasing phase angle, (iii) become moderate in the NIR, and (iv) may have a peak at $\lambda\sim1.6\ \micron$.

Similar trends were reported for Hale-Bopp, 1P/Halley, and West \citep{1978PASJ...30..161O, 1987A&A...187..621B, 1987A&A...187..689K, 1997EM&P...78..353H, 2006JQSRT.100..179K}. In our calculation, Hale-Bopp shows $c_{RH}=3.37\pm1.33\ \% / \micron$ and $c_{HK}=-6.73\pm2.87\ \% / \micron$ ($\alpha=48\degr$). 1P/Halley shows $c_{RH}=2.87\pm0.31\ \% / \micron$ and $c_{HK}=1.36\pm0.77\ \% / \micron$ ($\alpha=46\degr$), $c_{RH}=7.49\pm1.31\ \% / \micron$ and $c_{HK}=1.24\pm2.52\ \% / \micron$ ($\alpha=65\degr$), respectively. Finally, West shows $c_{RH}=5.89\pm0.82\ \% / \micron$ and $c_{HK}=-3.00\pm1.80\ \% / \micron$ ($\alpha=66\degr$), $c_{RH}=6.52\pm0.70\ \% / \micron$ and $c_{HK}=-22.17\pm1.20\ \% / \micron$ ($\alpha=\alpha_{\mathrm{max}}$), respectively. We make a detailed comparison of our results with these comets in Section \ref{sec:discussion}.

\section{Discussion}\label{sec:discussion}

In this section, we discuss the significance of our findings. Our polarimetric data of T-A are among the most complete data sets ever investigated for other comets in terms of its wide wavelength coverage (at 0.5--2.5 \micron) and phase angle range ($\alpha=$20--123\degr) and careful consideration of gas contamination.

In the optical wavelength, we derived reliable (i.e. less contaminated) polarimetric data that are supported by our spectroscopy. The optical PPC is roughly consistent with that of the majority of other comets except for a few comet samples. One of the outliers in PPC is Hale-Bopp \citep{1997EM&P...78..347K, 2000Icar..145..203M, 2006JQSRT.100..179K}. This comet is different from other comets in terms of its large nuclear size \citep[30--70 km in size;][]{1997EM&P...79...17W}. Such large icy bodies may indicate physical characteristics different from the majority of cometary nuclei ($\sim$1 km in size) due to thermal metamorphism by an internal heat source or geological processes \citep{2002EM&P...89...27P}. The high polarization of 2P/Encke is also different from most comets \citep{2005A&A...441..773J, 2018A&A...620A.161K}. Due to its small perihelion distance, refractories from 2P/Encke may be subject to intensive solar heating in proximity to the Sun \citep{2018A&A...620A.161K}. Excluding comets affected by these evolutionary processes, it could be argued that the polarization properties at the optical wavelength are largely consistent with each other.

Various theoretical light-scattering simulations of comet-like dust grains have shown that PPCs change dramatically depending on different optical constants, the size of the dust grain (including the size of the monomer in the dust aggregate), and the porosity \citep{2006A&A...449.1243K, 2009ApJ...696.2126S, 2010A&A...513A..40K}. Compared to the difference ($\gg$10 \%) predicted by the theoretical models with different assumptions, the observed polarimetric profiles (PPC and wavelength dependence) of T-A are roughly consistent ($\ll$10\%) with those of other comets. Accordingly, the physical properties (optical constants and grain size) of dust from OCCs (from the gas planet-forming region) and ECs (beyond the planet-forming region) are largely similar to each other. This interpretation further suggests that dust refractories beyond the snowline in the protoplanetary disk would be uniform (in terms of composition and monomer size) and would have experienced a similar coalescence process.

T-A's polarization wavelength dependence is close to Halley's but slightly higher overall than West's. However, there seems to be a common trend that polarimetric slope ($c_{i,j}$) decreases with increasing wavelength and takes nearly zero or negative values at wavelengths longer than 1.6 \micron\ to all three comets (see Figure \ref{fig:pcolor}). \citet{2024A&ARv..32....7B} suggested that this is due to depolarization by thermal radiation or packing density of dust aggregates. In particular, because West was observed at 0.43 au from the Sun, the depolarization of thermal radiation is quite significant. After the subtraction of the thermal radiation component, the $c_{i,j}$ value is almost zero at values above 1.6 \micron\ \citep{1978PASJ...30..161O}. We recognize it impossible to evaluate the effect of thermal radiation from our T-A data alone. However, we consider that the effect of thermal radiation would be negligibly small because the distance from the Sun is much larger than West. In the future, it is desirable to evaluate the contamination of thermal radiation if infrared data beyond the K-band is available. Therefore, the decrease in the value of $c_{i,j}$ above 1.6 \micron\ can be attributed to the porosity of the cometary dust.

\section{Summary} \label{sec:summary}

We present the contemporaneous polarimetric data of T-A with a wide coverage of phase angle (20--123\degr) and spectral range (0.5--2.5 \micron). It is one of the most complete polarimetric data sets among the comets. We confirmed that our polarimetric data are not contaminated by significant molecular gas emissions and possibly thermal emissions. Therefore, our data reflect only the physical properties of the cometary dust. The results show that (i) the values of $P_\mathrm{max}$ and $P_\mathrm{r}$ increase with wavelength but (ii) the trend moderates in the NIR, and (iii) may have a peak near the wavelength of 1.6 \micron. We conclude that PPCs and the spectral behavior of T-A are largely consistent with other comets regardless of the cometary origins (OCCs and ECs) in the protoplanetary disk. 

\begin{acknowledgments}
This work was supported by a National Research Foundation of Korea (NRF) grant funded by the Korean government (MEST) (No.2023R1A2C1006180).
\end{acknowledgments}

\appendix

\onecolumngrid
\section{Aperture polarimetry}\label{appendix:appol}

\setcounter{table}{0}
\renewcommand{\thetable}{A\arabic{table}}
\setcounter{figure}{0}
\renewcommand{\thefigure}{A\arabic{figure}}

In this section, we summarize our data analysis procedure for aperture polarimetry in SQUIDPOL (Rc and Ic) and NIC (J, H, and Ks). The preprocessing includes conventional bias, dark, and flat corrections, along with cosmic-ray rejection. In case of NIC, we also corrected the vertical pattern noise from the raw images using the NIC data reduction package (NICpolpy, \citealt{2022StGal...5....4B}). After preprocessing, we extracted the source flux from ordinary and extraordinary components as described in Section \ref{subsec:phaseangle}. 

We obtained the linear polarization degree ($P$) and the electric vector position angle ($\theta_P$) using

\begin{equation}\label{eq:poldegree}
    P=\sqrt{(Q/I)^2+(U/I)^2}
    \quad\mathrm{and}\quad 
    \theta_P=\frac{1}{2}\tan^{-1}{(\frac{U}{Q})} ~,
\end{equation}

\noindent where $Q$, $U$, and $I$ are the Stokes parameters of the extracted flux. Before adapting Equation (\ref{eq:poldegree}), we calibrated the Stokes parameters with instrumental polarization ($q_\mathrm{inst}$ and $u_\mathrm{inst}$), polarization efficiency ($p_\mathrm{eff}$), and position angle offset using the calibration parameters given in Table \ref{tab:polparams}. In the case of SQUIDPOL, the $p_\mathrm{eff}$ values have not yet been established; thus we assumed $p_\mathrm{eff}=1$. Although the precise values of $p_\mathrm{eff}$ will be provided in a future study, this assumption is justified since laboratory experiments and preliminary observational results have confirmed that $p_\mathrm{eff}$ exceeds 90 \% for Rc- and Ic-bands. After polarimetric calibration, we calculated the weighted mean of the normalized parameters ($Q/I$ and $U/I$) and derived the polarimetric result. The errors of $P$ and $\theta_P$ ($\sigma_P$ and $\sigma_{\theta_P}$) were derived by error propagation based on random errors in the measured Stokes parameters.

To correct for the inherent bias of observed $P$ due to its random noise, we corrected the resultant polarization degree with the following equation \citep{1974ApJ...194..249W}:

\begin{equation}\label{eq:poldebias}
    P_\mathrm{debias}=\sqrt{P^2-\sigma^2_P}~.
\end{equation}

When $P < \sigma_P$ the above equation raises mathematical errors, and we treated $P=0\ \%$ in this case (In our dataset, this is the case for Rc-band data in 2024 December 17).

Finally, we obtained the polarimetric results relative to the scattering plane using

\begin{equation}\label{eq:polscattering}
    P_\mathrm{r}=P_\mathrm{debias}\cos{(2\theta_\mathrm{r})}
    \quad\mathrm{and}\quad 
    \theta_\mathrm{r}=\theta_{P}-(\phi \pm 90\degr) ~,
\end{equation}

\noindent where $\phi$ is the position angle of the scattering plane at the time of observation. We selected one of the $\pm$ signs which satisfy $-90\degr<\theta_\mathrm{r}<90\degr$.

Table \ref{tab:impol} summarizes the final results of aperture polarimetry in the five bands. 

\begin{deluxetable*}{cccccccc}[!htbp]
\tablecaption{Calibration parameters in aperture polarimetry. \label{tab:polparams}}
\tablehead{\colhead{Instrument} & \colhead{Filter} & \colhead{$\lambda_\mathrm{c}$ [\micron]} & \colhead{$\lambda_\mathrm{fwhm}$ [\micron]} & \colhead{$q_\mathrm{inst}$ [\%]} & \colhead{$u_\mathrm{inst}$ [\%]} & \colhead{$p_\mathrm{eff}$ [\%]} & \colhead{Position ang. offset [\degr]} \\
& & (1) & (2) & (3) & (4) & (5) & (6)}
\startdata
SQUIDPOL & Rc & 0.64 & 0.16 & $-0.04\pm0.19$ & $-0.08\pm0.20$ & - & $4.2\pm2.3$ \\
         & Ic & 0.80 & 0.15 & $-1.53\pm0.19$ & $0.15\pm0.14$ & - & $8.5\pm1.9$ \\
\hline
NIC      & J  & 1.25 & 0.29 & $-0.00\pm0.29$ & $-0.01\pm0.29$ & $98\pm6$  & $0.5\pm1.3$ \\
         & H  & 1.65 & 0.30 & $0.03\pm0.52$  & $-0.03\pm0.55$ & $95\pm7$  & $1.3\pm3.1$ \\
         & Ks & 2.15 & 0.32 & $-0.02\pm0.30$ & $-0.07\pm0.31$ & $92\pm12$ & $-0.7\pm6.3$ \\
\enddata
\tablecomments{(1) Central wavelength of filters. (2) Bandwidth of filters. (3) Instrumental polarization ($Q/I$). (4) Instrumental polarization ($U/I$). (5) Polarization efficiency. (6) Position angle offset. Parameters of NIC are reproduced from \citet{2019StGal...2....3T}. The polarization efficiencies of SQUIDPOL have not yet been established.}
\end{deluxetable*}

\begin{deluxetable*}{cccccccccc}[!htbp]
\tablecaption{Summary of aperture polarimetry. \label{tab:impol}}
\tablehead{
\colhead{UT Date} & \colhead{$\alpha$ [\degr]} & \colhead{Filter} & \colhead{$P$ [\%]} & \colhead{$\sigma_P$ [\%]} & \colhead{$\theta_P$ [\degr]} & \colhead{$\sigma_{\theta_P}$ [\degr]} & \colhead{$P_r$ [\%]} & \colhead{$\theta_r$ [\degr]}\\
& & & (1) & (2) & (3) & (4) & (5) & (6)}
\startdata
\hline
2024-10-16 & 123.15 & Rc & 21.78 & 0.73 & 162.70 & 0.96 & 21.10 & 1.75 \\
 &  & Ic & 22.98 & 1.10 & 164.61 & 1.37 & 22.22 & 3.66 \\
\hline
2024-10-17 & 116.48 & Rc & 24.94 & 1.01 & 163.45 & 1.16 & 24.15 & 1.93 \\
 &  & Ic & 25.72 & 1.78 & 164.70 & 1.98 & 24.92 & 3.18 \\
\hline
2024-10-20 & 99.68 & Rc & 28.80 & 2.29 & 163.30 & 2.28 & 27.93 & 1.08 \\
 &  & Ic & 31.43 & 5.04 & 164.75 & 4.60 & 30.53 & 2.53 \\
\hline
2024-10-23 & 86.71 & Rc & 28.48 & 0.79 & 163.33 & 0.79 & 30.74 & 1.22 \\
 &  & Ic & 29.90 & 0.79 & 164.35 & 0.76 & 32.79 & 2.24 \\
 &  & J & 34.37 & 0.95 & -69.88 & 0.79 & 34.27 & 2.23 \\
 &  & H & 35.81 & 2.98 & -69.71 & 2.38 & 35.68 & 2.41 \\
 &  & Ks & 34.39 & 2.52 & -68.60 & 2.10 & 34.13 & 3.51 \\
\hline
2024-10-24 & 82.98 & Rc & 28.74 & 0.79 & 163.37 & 0.78 & 31.01 & 1.40 \\
 &  & Ic & 30.34 & 0.70 & 164.55 & 0.66 & 33.24 & 2.58 \\
\hline
2024-10-25 & 79.78 & J & 32.73 & 2.06 & -76.22 & 1.81 & 32.34 & -4.43 \\
 &  & H & 35.10 & 3.02 & -77.13 & 2.47 & 34.49 & -5.34 \\
 &  & Ks & 34.22 & 5.21 & -72.27 & 4.36 & 34.22 & -0.48 \\
\hline
2024-10-30 & 66.22 & Rc & 20.14 & 0.71 & 161.82 & 1.02 & 21.72 & 1.41 \\
 &  & Ic & 23.29 & 0.66 & 161.97 & 0.81 & 25.58 & 1.56 \\
 &  & J & 25.62 & 0.10 & -67.59 & 0.11 & 25.50 & 2.82 \\
 &  & H & 26.76 & 0.09 & -67.94 & 0.10 & 26.66 & 2.48 \\
 &  & Ks & 27.12 & 0.10 & -66.94 & 0.11 & 26.92 & 3.47 \\
\hline
2024-10-31 & 63.96 & Rc & 17.68 & 0.70 & 161.63 & 1.13 & 19.07 & 1.56 \\
 &  & Ic & 20.59 & 0.65 & 162.76 & 0.91 & 22.60 & 2.69 \\
\hline
2024-11-11 & 46.46 & J & 13.26 & 0.09 & -62.96 & 0.18 & 13.21 & 2.52 \\
 &  & H & 14.21 & 0.08 & -63.33 & 0.16 & 14.17 & 2.14 \\
 &  & Ks & 15.57 & 0.18 & -61.47 & 0.34 & 15.42 & 4.00 \\
\hline
2024-11-12 & 45.28 & Rc & 10.91 & 0.67 & 161.40 & 1.77 & 11.49 & 6.41 \\
 &  & Ic & 13.04 & 0.63 & 163.41 & 1.39 & 13.73 & 8.42 \\
 &  & J & 12.77 & 0.08 & -62.71 & 0.18 & 12.73 & 2.29 \\
 &  & H & 13.57 & 0.07 & -63.04 & 0.14 & 13.54 & 1.96 \\
 &  & Ks & 15.16 & 0.10 & -61.61 & 0.18 & 15.06 & 3.39 \\
\hline
2024-11-19 & 38.08 & Rc & 7.04 & 0.68 & 152.85 & 2.77 & 7.59 & 1.48 \\
\hline
2024-12-17 & 20.61 & Rc & 0.00 & 0.65 & - & - & 0.00 & - \\
\enddata
\tablecomments{(1) Observed polarization degree. (2) Standard deviation of $P$. (3) Electric vector position angle. (4) Standard deviation of $\theta_P$. (5) Polarization degree relative to the scattering plane. (6) Electric vector position angle relative to the scattering plane.}
\end{deluxetable*}

\onecolumngrid
\section{Spectra on different nights}\label{appendix:spectrum}

\setcounter{table}{0}
\renewcommand{\thetable}{B\arabic{table}}
\setcounter{figure}{0}
\renewcommand{\thefigure}{B\arabic{figure}}

\begin{figure*}[!htbp]
    \centering
    \includegraphics[width=1.0\linewidth]{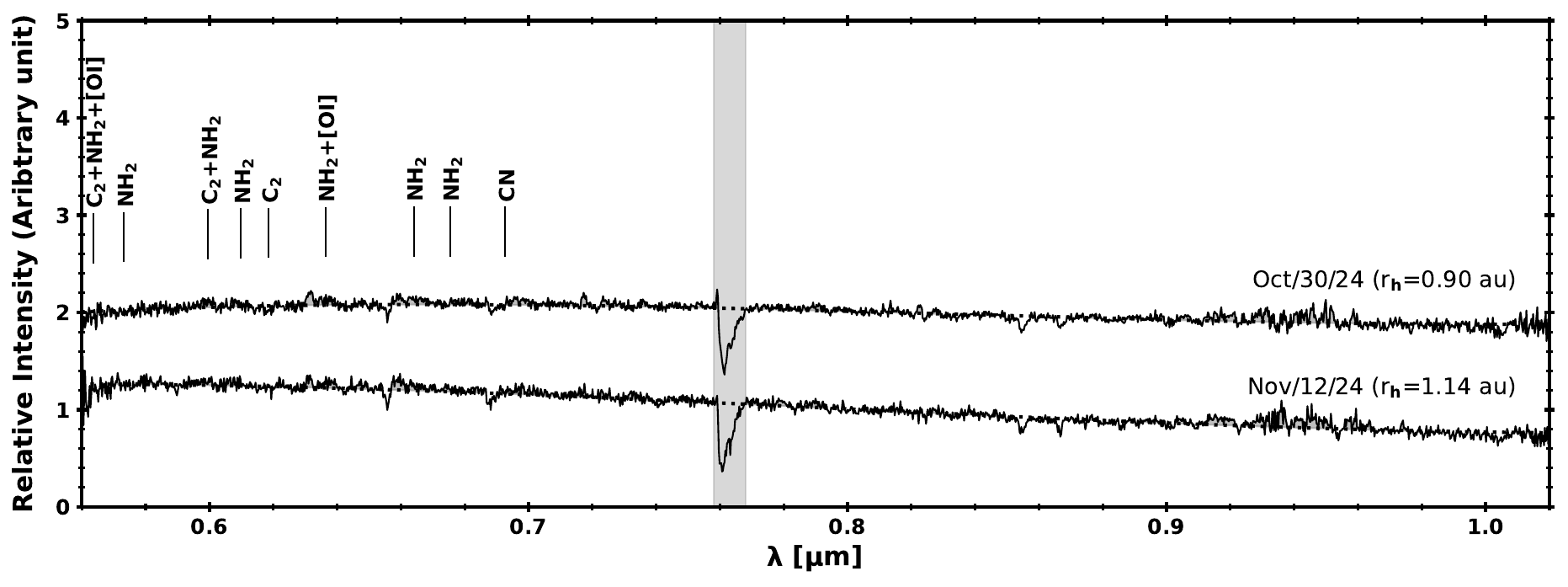}
    \caption{Same with Figure \ref{fig:spectrum}, but showing the spectrum on two different nights (October 30 and November 12, 2024). The spectrum is normalized with the flux at $\lambda=0.8\ \micron$, while adding unity on Oct 30 for visibility. The negative spikes near $\lambda=0.76\ \micron$ are the artifacts during the flux calibration due to the imperfect correction of telluric absorption.}
    \label{appendixfig:spectrum_01}
\end{figure*}

Figure \ref{appendixfig:spectrum_01} presents the optical spectrum of T-A ($\lambda=0.56$--$1.00\ \micron$) on two nights (October 30 and November 12). The data were taken using HONIR sppol mode, bias, dark, flat corrected, median-combined, and flux-calibrated using standard stars with airmass ($X$) similar to T-A ($\Delta X < 0.1$) at the time of observation. (We did not apply any airmass correction for the standard star’s spectral slope.) The slit position angle was fixed to the North-South direction to avoid contamination from the dust tail, which extended toward the Northeast during the observation period. 

\begin{figure*}[!htbp]
    \centering
    \includegraphics[width=1.0\linewidth]{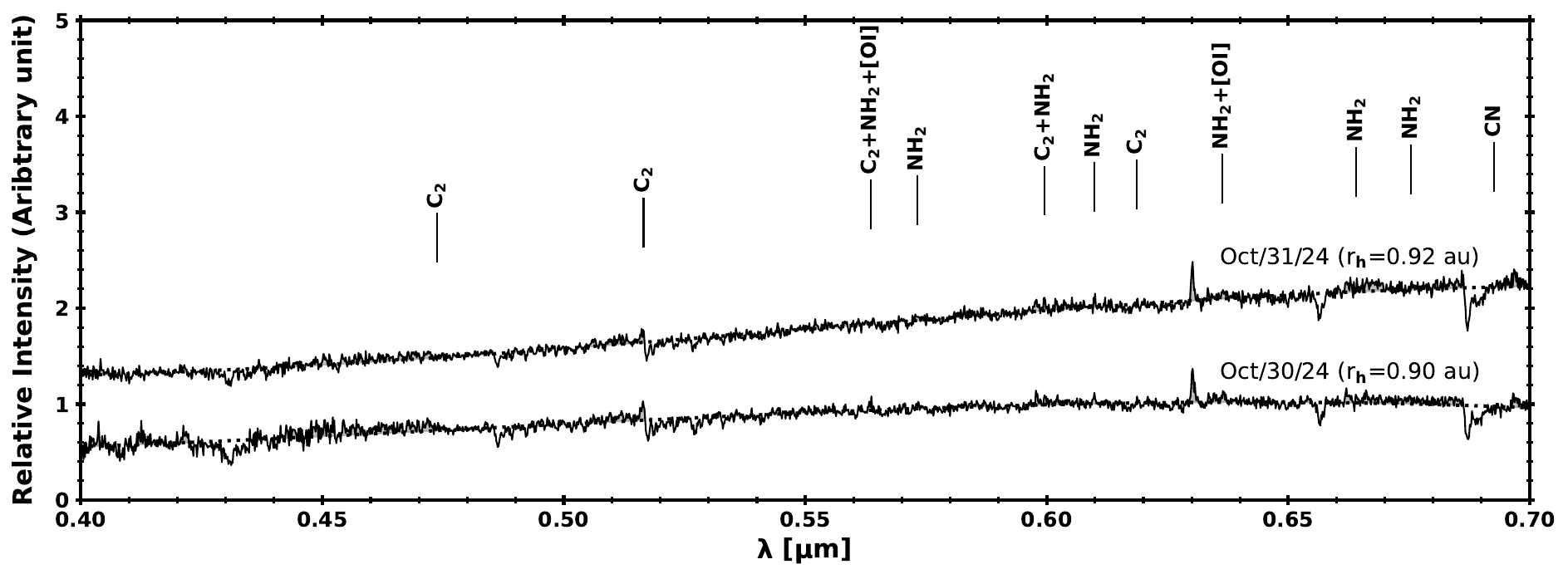}
    \caption{The spectrum on two different nights (October 30 and 31, 2024) taken with LISA spectrograph. The spectrum is normalized with the flux at $\lambda=0.6\ \micron$, while adding unity on Oct 31 for visibility.}
    \label{appendixfig:spectrum_02}
\end{figure*}

Figure \ref{appendixfig:spectrum_02} presents the optical spectrum of T-A ($\lambda=0.4$--$0.7\ \micron$) on two nights (October 30 and 31). The data were taken using LISA spectrograph, bias, dark, flat corrected, median-combined, and flux-calibrated using standard stars with airmass similar to T-A ($\Delta X<0.2$) at the time of observation. Before flux calibration, we corrected the standard star spectrum for airmass effects. The slit position angle was fixed to the North-South direction to avoid contamination from the dust tail, which extended toward the Northeast during the observation period.

\clearpage

\onecolumngrid
\section{Spectropolarimetry}\label{appendix:sppol}
\setcounter{table}{0}
\renewcommand{\thetable}{C\arabic{table}}
\setcounter{figure}{0}
\renewcommand{\thefigure}{C\arabic{figure}}

\begin{figure*}[!htbp]
    \centering
    \includegraphics[width=1.0\linewidth]{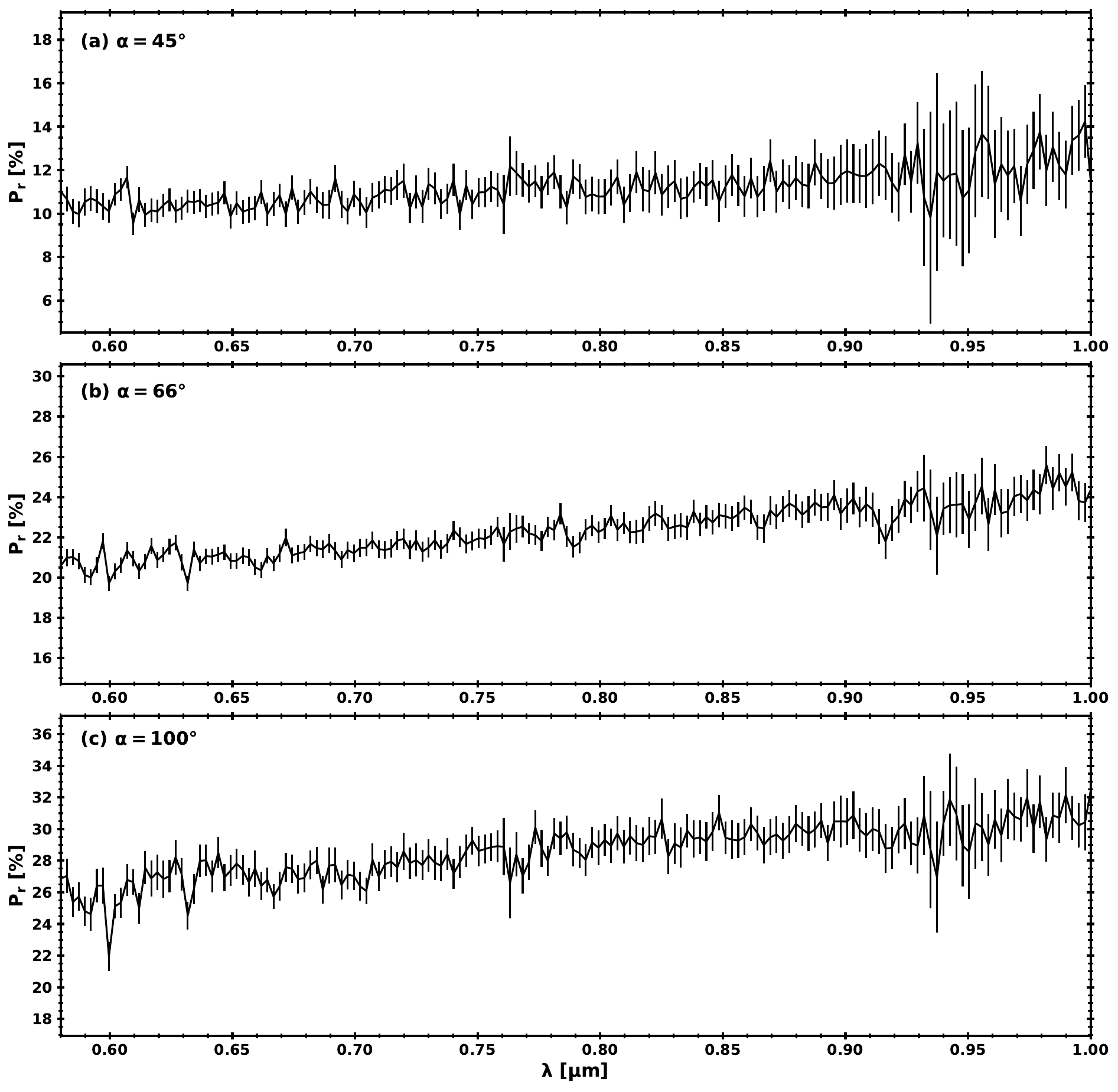}
    \caption{Spectropolarimetry of T-A at phase angles (a) $\alpha=45\degr$, (b) $\alpha=66\degr$, and (c) $\alpha=100\degr$.}
    \label{appendixfig:sppol}
\end{figure*}

Figure \ref{appendixfig:sppol} presents the spectropolarimetry of T-A on three nights (Oct 20, 30, and November 12). The ordinary and extraordinary components were taken using HONIR sppol mode, then bias, dark, and flat were corrected. The overall polarimetric result was obtained similarly to that in the process of aperture polarimetry (Appendix \ref{appendix:appol}), while the calibration parameter is given by \citet{2014SPIE.9147E..4OA}. The instrumental polarization of HONIR in optical is $<0.2\ \%$.

\end{document}